\documentclass[11pt,twocolumn]{aastex61}

\usepackage{gensymb}
\usepackage{amsmath}
\usepackage{bm}
\usepackage{rotating}
\usepackage{longtable}

\newcommand{\HI}{\mbox{H\,{\sc i}}}

\newcommand{\Htwo}{H$_{2}$}

\newcommand{\twCO}{$^{12}$CO}
\newcommand{\thCO}{$^{13}$CO}

\newcommand{\kmps}{\mbox{${\rm km\;s^{-1}}$}}

\newcommand{\lsim}{\mbox{$\mathrel{\vcenter{\hbox{\ooalign{\raise3pt\hbox{$<$}\crcr \lower3pt\hbox{$\sim$}}}}}$}}
\newcommand{\gsim}{\mbox{$\mathrel{\vcenter{\hbox{\ooalign{\raise3pt\hbox{$>$}\crcr \lower3pt\hbox{$\sim$}}}}}$}}
\shorttitle{Draft Paper on W5 OH Column Densities}
\shortauthors{Philip Engelke, Ronald J. Allen}

\begin{document}


\title{
OH as an Alternate Tracer for Molecular Gas: \\ Quantity and Structure of Molecular Gas in W5}

\correspondingauthor{Philip Engelke}

\author[0000-0002-0786-7307]{Philip D. Engelke}
\altaffiliation{Grote Reber Fellow}
\affiliation{Department of Physics and Astronomy, The Johns Hopkins University, Baltimore, MD 21218, USA}
\affiliation{National Radio Astronomy Observatory, 1003 Lopezville Road, Socorro, NM 87801, USA}
\email{pengelk1@jhu.edu}

\author[0000-0001-9906-8352]{Ronald J. Allen}
\affiliation{Space Telescope Science Institute, 3700 San Martin Drive, Baltimore, MD 21218, USA}
\affiliation{Center for Astrophysical Sciences, Department of Physics and Astronomy, The Johns Hopkins University, Baltimore, MD 21218, USA}
\email{rjallen@stsci.edu}

\begin{abstract}

We report column densities of molecular gas in the W5 star-forming region as traced with OH 18-cm emission in a grid survey using the Green Bank Telescope. OH appears to trace a greater column density than does CO in 8 out of 15 cases containing OH emission detections; the two molecules trace the same column densities for the other 7 cases. OH and CO trace a similar morphology of molecular gas with a nearly one-to-one correspondence. The mass of molecular gas traced by OH in the portion of the survey containing OH emission is $1.7$ (+ 0.6 or - 0.2) $\times 10^4 M_{\odot}$, whereas the corresponding CO detections trace $9.9 \times 10^3 M_{\odot} (\pm 0.7) \times 10^3$. We find that for lines observed in absorption, calculations assuming uniform gas and continuum distributions underestimate column density values by 1 to 2 orders of magnitude, making them unreliable for our purposes. Modeling of this behavior in terms of OH cloud structure on a scale smaller than telescopic resolution leads us to estimate that the filling factor of OH gas is a few to 10 percent. Consideration of filling factor effects also results in a method of constraining the excitation temperature values. The total molecular gas content of W5 may be approximately two to three times what we report from direct measurement, because we excluded absorption line detections from the mass estimate.


\end{abstract}

\section{Introduction}


\subsection{Background}

 
Molecular gas is an important part of the interstellar medium (ISM), but its primary component \Htwo\ is difficult to detect in clouds of diffuse gas owing to insufficient collisional excitation of the lowest rotational levels and the lack of a dipole moment. As a result, radio signals from asymmetric  molecules with larger moments of inertia are used as surrogate tracers for \Htwo.  Since its discovery in 1970 \cite[]{wjp70}, the (1-0) transition of \twCO\ near 115 GHz is almost universally used as a tracer of the large-scale distribution of molecular gas in our Galaxy as well as external galaxies, and has produced a wealth of knowledge about interstellar molecular gas.

However, over the past several decades, numerous questions have persisted concerning the use of the \twCO(1-0) line for accurate quantitative work. First, CO may not be an ideal tracer for measuring molecular gas below the critical density of $\sim 10^3$ cm$^{-3}/\tau$ for this transition, where $\tau$ is the optical depth of the gas. Additionally, when a CO(1-0) line is detected, its high optical thickness requires use of an indirect method (the ``X-factor'' method) to calculate column densities instead of a direct calculation using the equation of radiative transfer. The exact value of this ``X-factor'' is uncertain, and variations of up to an order of magnitude have been reported \citep[see][for a review]{bwl13}. Additional information about CO content can be gleaned from other sources such as \thCO(1-0), when such data are available. Finally, recent gamma-ray and IR surveys have provided evidence of undetected gas containing hydrogen nuclei in the Galaxy; this gas is believed to be molecular, and may contain a mass comparable to the currently known mass of molecular gas in the Galaxy \cite[][]{gct05, abdo10, t15}. Evidence for undetected gas is also provided by dust extinction and emission studies \cite[e.g.][]{paradis12, lee15}, and from C$^+$ emission \cite{pineda13, langer14, tang16}. Unfortunately, the specific nature of this ``CO-dark'' gas is not provided by the gamma-ray and IR studies, nor is kinematic information or accurate structure and mass information available.

An alternative to \twCO(1-0) as a tracer for the large-scale distribution and kinematics of diffuse interstellar gas is thus desirable. In this paper we continue to explore the use of the 18-cm lines of the OH molecule as one  alternative tracer. These lines are optically thin, have low critical density ($\sim 10$ cm$^{-3}$), and have been widely detected in the ISM when observations of sufficient sensitivity are made \citep[see ][for early results and a brief historical review]{arb12, arb13}. The SPLASH survey \cite[]{dawson14} using the Parkes telescope pointed towards the inner Galaxy also found widespread OH at all four 18-cm lines, in both emission and absorption. \cite{ahe15} used the GBT (FWHM $\approx 7.6'$) to carry out a ``blind'' survey of 18-cm OH emission on a coarse grid of positions in the direction of a quiescent region in the outer Galaxy, with two hours of exposure at each coordinate. Widespread OH emission was detected from both main lines at 1665 and 1667 MHz as well as the 1720 satellite line, though observations of the 1612 satellite line of OH were compromised by radio-frequency interference. Within the sensitivity limit of the observations, the intensities of the two main lines were overwhelmingly in the ratio 5:9 as expected for LTE conditions. Compared to the \twCO(1-0) data from the CfA survey \cite[]{dht01}, many OH features were identified that did not show any corresponding CO emission. All CO features, however, did have corresponding OH emission. A plot of profile integrals of the \twCO\ features versus those of the corresponding OH features reveals a ``bimodal'' distribution, with approximately half of the pointings showing little or no \twCO\ emission, and the remainder displaying a rough correlation between the strengths of the OH and \twCO\ lines. The conclusion was that OH traces a significantly larger component of the molecular ISM than does \twCO.

These results suggest that OH observations may provide new insight in other situations where more precise estimates of the quantity and kinematics of molecular gas could shed light on large-scale astrophysical processes. In this paper we use OH as a tool to measure the quantity of \Htwo\ in close proximity to a star forming region, and compare that result with the quantity of \Htwo\ estimated using \twCO\ emission and the X-Factor.

\subsection{The W5 Star-forming Region}

We chose W5 as a star-forming region to study with a GBT grid survey. As first mentioned in \cite{ea18}, the reasons for this choice are as follows. First, W5 covers an angular size on the sky that is large enough for a grid survey with the GBT to reveal sufficient structural detail while also not requiring an inappropriately-long observing time. Second, there are several radio continuum maps in the literature covering the region, including the Canadian Galactic Plane Survey at 408 and 1420 MHz \cite[the CGPS;][]{tgp03}, and a 2695 MHz survey with the Effelsberg telescope \cite[]{frrr90}. A radio continuum study focusing on the W3-W4-W5 complex associated with the CGPS is also available \cite[]{norm97}. These surveys show that the continuum temperatures associated with W5 are in the range of excitation temperatures typical for OH in the general ISM, implying that OH in front of the nebula may appear either in emission or in absorption. This fortunate situation has permitted an accurate measurement of the excitation temperatures for both the 1665 and 1667 OH lines in close proximity to this star-forming region \citep[][]{ea18}. Finally, W5 is relatively simple in structure; it does not contain large numbers of stars or abundant IR radiation (which is known to lead to enhanced OH emission, complicating the calculation of OH column density) in comparison to other star forming regions \cite[]{km03,ml05} . While minor peculiar motion is still present, the radial velocity ranges are not large enough to make it difficult to identify W5 features by a kinematic velocity range: -49 km/s to -31 km/s, which we adopt from \cite{km03}. 

In order to use an OH survey of W5 to study structure and mass of molecular gas clouds in the region, it is necessary to know the distance to W5. Finding the distance to W5 is not as simple as using kinematic distance calculations based on Galactic rotation models, as anomalous velocities are also involved \cite[]{xu06,dza12}. OH maser parallax measurements using the Very Long Baseline Array (VLBA) reveal the distance to W3(OH), an OH maser feature within W3, to be $2.00 \pm 0.05$ kpc \cite[]{xu06,hach06}, as mentioned by \cite{dza12}. This measurement could provide a distance to the entire W3-W4-W5 complex if we assume that all three HII regions are a correlated group and are the same distance away. As discussed by \cite{dza12}, distances to W4 and W5 have been provided by spectrophotometric measurements of OB clusters IC 1805 in W4 and IC 1848 in W5, both of which are 2.1 to 2.4 kpc \cite[]{bf71,m72,m95,chau09,chau11}. These measurements are independent of the VLBI measurement of the distance to W3 and suggest that W3, W4, and W5 are all in close proximity and at the same distance. \citet[][]{dza12} suggest using a value of 2.0 kpc for the distance to W5, which we adopt for this paper. 


\section{Observing Program}

\subsection{The Survey}

\begin{figure*}[ht!] 
\epsscale{0.9}
\vspace{0.1in}
\begin{center}
\includegraphics[width=0.8\textwidth,angle=0]{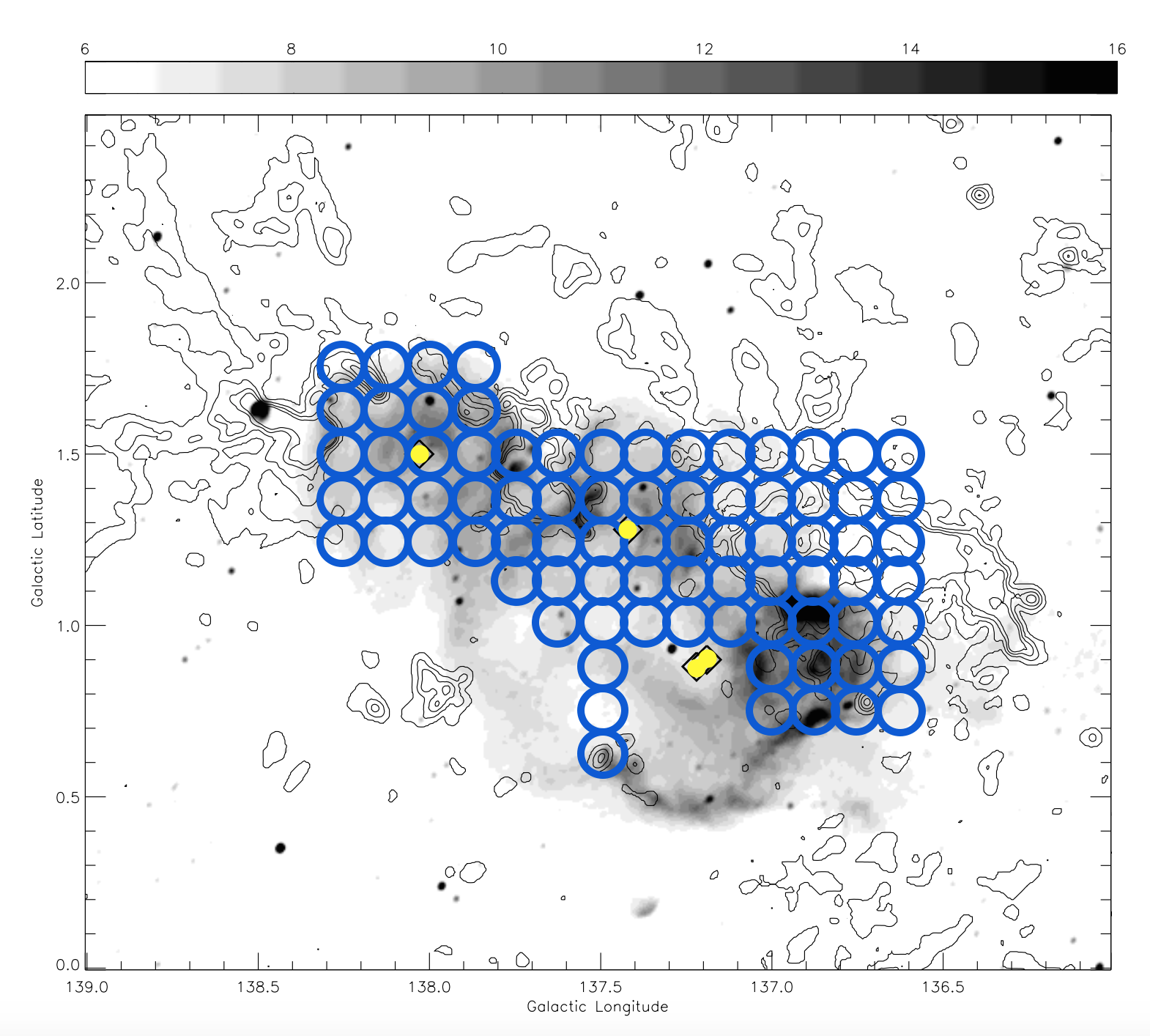} \hspace{0.2in} 
\caption{Positions of our GBT observations overlaid on the map of the W5 region taken directly from \citet[][Figure 2]{km03}. The gray scale indicates the 1420 MHz continuum from the CGPS survey \cite[]{tgp03}, the contours indicate \twCO(1-0) emission from the FCRAO survey \cite[]{hbs98} integrated within the LSR radial velocity range of -49 to -31 \kmps, and the 4 diamonds (which we have emphasized in yellow) indicate the locations of known O stars in W5 as plotted by \cite{km03}. We have added a number of blue circles indicating the positions of the GBT pointings in our grid survey; the circle size indicates the $\approx 7\farcm6$ angular resolution FWHM of the GBT. \label{fig:SurveyMap}} 
\end{center}
\end{figure*}

The goal for the design of our OH survey was to obtain OH column density results from a major part of any molecular clouds located in the direction of W5. In order to study the structure of W5 with sufficient detail while keeping the total observing time within reasonable limits, we chose a representative set of 80 positions covering a large fraction of the radio continuum image of the region. The choice of the specific pointings to observe was primarily informed by the brightness of the radio continuum and by the distribution of \twCO(1-0) emission from the FCRAO survey \cite[]{hbs98}. Figure \ref{fig:SurveyMap} shows the observed positions in our survey superimposed on a map taken from \citet[][their Figure 2]{km03} that displays the 1420 MHz continuum emission in gray scale and the integrated CO content in contours. 



\subsection{Observations}
The observations for this work were made with the 100-m Robert C. Byrd Green Bank Telescope (GBT) in West Virginia. OH spectra were acquired with the VEGAS spectrometer in the L-band, with 2 hours of exposure per pointing. The system temperature was typically less than 20 K. We used frequency-switching mode because the ubiquitous nature of faint OH emission in the sky means that it is not always possible to find a nearby reference position for position switching \cite[see][]{ahe15}. We observed all four OH 18-cm transitions at 1612, 1665, 1667, and 1720 MHz, plus the 1420 MHz HI line in dual polarization. The spectrum at each IF band was approximately 12.8 MHz wide, so we were able to observe the 1665 and 1667 MHz lines simultaneously in a single IF band. The FWHM of the GBT point spread function at 1667 MHz is 7\farcm6, and the channel step is 0.515 kHz corresponding to 0.0926 \kmps\ at 1667 MHz. Each 2-hour pointing includes 12 ``scans'' each of 10 minutes integration. A total of 80 GBT pointings was observed between February 2016 and April 2017 in GBT programs 16A-354 and 17A-386. The maximum spectral dump rate was every 0.5 ms. Radio frequency interference was largely insignificant for the majority of scans at the observed frequencies except for 1612 MHz, which limited the utility of the data at that frequency.

\subsection{Data Reduction}

For data reduction of the two main OH lines near 1666 MHz, we averaged the twelve ten-minute scans for each pointing separately for XX and YY polarizations, excluding any scans containing RFI contamination. Then we averaged together the two polarizations. There is a large-scale baseline ripple of frequency 9 MHz, which is an instrumental effect of the GBT appearing at high sensitivity levels. We fit this baseline with a fifth-degree polynomial in the frequency range 1661.4 MHz to 1671.4 MHz and subtracted the fit from the entire spectrum. This baseline fit does not alter the OH spectral lines themselves because the scale of the baseline ripple is so much larger than the scale of any OH spectral features, and the baseline in the vicinity of spectral features will be defined as zero, as we describe later. Next, we identified the 1665 and 1667 MHz portions of the spectrum and saved the data separately for the two main line frequencies. 

Each of the two main lines was then subjected to a final fit to a linear baseline over the immediate velocity range of any OH features visible in the spectrum.
For an independent estimate of the velocity range boundaries of any spectral features in W5, we consulted the \HI\ profiles from the Dominion Radio Astrophysical Observatory (DRAO) low resolution survey \cite[]{ht00,higgs05} in addition to a visual inspection of the OH spectrum. The velocity range of these baseline fits is approximately 10 \kmps\ beyond the left and right edges of the spectral feature. Providing a good fit to the rest of the baseline is not necessary because we are not concerned with parts of the spectrum located far from the W5 velocity range. Finally, we calculated the strength of each line by summing the signal in spectrometer channels covering the spectral line within the radial velocity range -49 to -31 \kmps\ . 
%
%
The OH observations are of high sensitivity, with typical RMS noise levels of $\sim$ 3 mK, and typical line detections are of $\sim 7\sigma$.

\begin{figure*}[ht!] 
\epsscale{0.9}
\vspace{0.1in}
\begin{center}
\includegraphics[width=0.7\textwidth,angle=0]{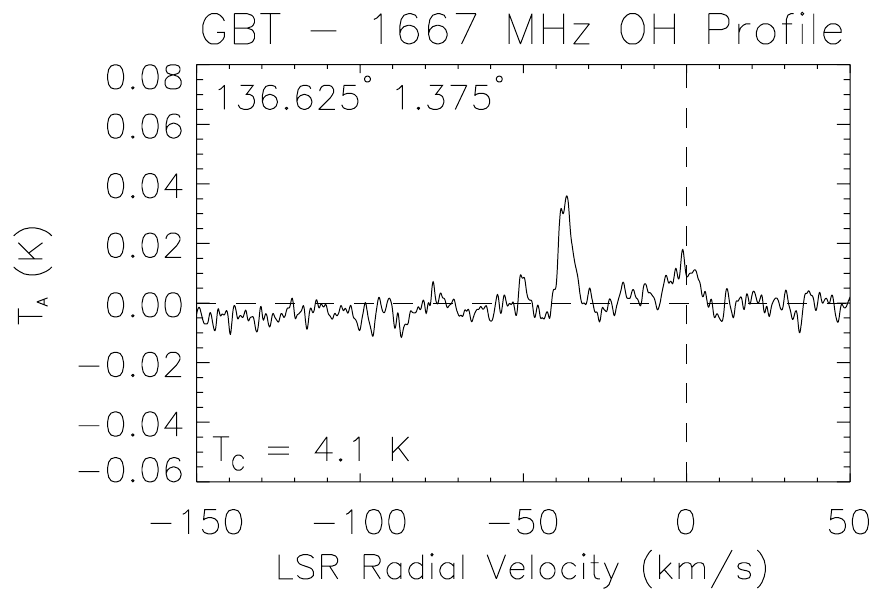} \hspace{0.2in} 
\caption{Example of a 1667 MHz OH spectrum from the W5 survey. This spectrum was taken at l = $136.625^{\circ}$, b = $1.375^{\circ}$ with 2 hours of exposure time on the GBT. We consider any feature detected between -49 km/s and -31 km/s to be associated with W5. In this spectrum, the OH feature associated with W5 is located near -38 km/s. Regions of the baseline measuring 10 km/s on the immediate left and right of the OH feature associated with W5 are fitted to zero. }\label{fig:SpectrumExample} 
\end{center}
\end{figure*}

\section{Calculating Column Densities}
\label{sec:coldens}
\subsection{OH Column Density Equations}

For emission lines, the following equation relates the OH column density to the observed line profile integral $\int{T_B(\nu)d\nu}$, continuum temperature $T_C$, and excitation temperature $T_{ex}$, where temperatures are measured in Kelvins \cite[see e.g. ][]{ll96}:

\begin{equation}
N(OH) = C\frac{T_{ex}}{T_{ex} - T_C}\int{T_b(\nu) d\nu}\textbf{,}
\label{eq:emission}
\end{equation}
where the coefficient $C_{67} = 2.3 \times 10^{14}$ for 1667 MHz if the resulting N(OH) is to be in terms of cm$^{-2}$, and $C_{65} = 4.1 \times 10^{14}$ for 1665 MHz, which is 9/5 times the 1667 MHz coefficient.

If the OH lines are observed in absorption, the following equation can be used, where $\tau(\nu)$ is the line optical depth $\tau(\nu) = ln(T_C/(T_C + T_b(\nu))$ \cite[see e.g. ][]{ll96}: 

\begin{equation}
N(OH) = C T_{ex} \int{\tau(\nu) d\nu}.
\label{eq:absorption}
\end{equation}

As noted in \cite{ea18}, the majority of spectra from the GBT survey do not show evidence of enhanced OH emission; deviations do occur from the 9:5 main line ratio, but most of these deviations can be explained entirely as a result of the difference in excitation temperatures for the two main lines as measured in \cite{ea18}. All emission features used in our column density analyses meet that criterion. 

The GBT observations in the grid survey directly provide the line profile integral for any detected OH signals. When both main line frequencies are detected in emission, we take the mean of the profile integrals to calculate N(OH). However, we must find $T_C$ and $T_{ex}$ by other means.

Since the GBT observations do not provide the continuum temperature, we must calculate what the $T_C$ value would be at 1667 MHz at the position of each GBT observation for a point spread function equivalent to that of the GBT observations. This process is described in \cite{ea18}. We calculate $T_C$ as would be observed at 1667 MHz with a point spread function like that of the GBT observations by interpolating radio continuum survey data at other frequencies \cite[]{frrr90,tgp03}, smoothed to the resolution of the GBT observations. The values of $T_C$ specifically located behind the OH gas are made more precise in certain cases (see Section \ref{sec:Filling Factor}).

Determining the excitation temperatures is more complicated and our use of the ``continuum background method'' to measure $T_{ex}$ from our survey data and $T_C$ values is described in detail in \cite{ea18} with further constraints discussed in Appendix \ref{app:Tex}. We note that in \cite{ea18}, we found that the main line excitation temperatures have two distinct values, and that it is important in doing column density calculations to use the appropriate values for the corresponding lines.

Details about the calculations and uncertainties in the column density determinations are provided in Section \ref{sec:Calculations and Uncertainties}.

\subsection{Total Molecular Gas Column Densities}

\label{sec:total}

In order to calculate N(\Htwo) as traced by OH, we start with our results for N(OH) and then multiply by the OH to H$_2$ conversion factor based on the abundance ratio for the molecules. \cite{wes10} report an abundance ratio of $1.05 \pm 0.14 \times 10^{-7}$ determined using a combination of archival and new measurements of UV absorption from OH and H$_2$ in front of several O and B stars, which we adopt in this paper. This result is also consistent with the value of $1.0 \pm 0.2 \times 10^{-7}$ reported by \cite{ll02}, $\sim 1.3 \times 10^{-7}$, from \cite{rugel18}, and $\sim 1.0 \times 10^{-7}$ from \cite{nguyen18}. A thorough summary of the OH to H$_2$ ratio is provided by \cite{nguyen18}, who also report that molecular gas content derived from dust reddening from \cite{green18} provides a more accurate result than that derived from infra-red data \cite[]{Planck}. We follow the dust reddening method of \cite{nguyen18} for measuring the OH to H$_2$ ratio on our own OH observations as an exercise to test the consistency the ratio in W5 with the values from elsewhere in the literature, using our HI measurements, OH measurements, and dust reddening from \cite{green18} as a proxy for total hydrogen nuclei to determine the OH to H$_2$ ratio. Although our optically thick HI data introduce sufficient uncertainty in the results as to deem them less useful than the \cite{wes10} value, we do show that OH abundances in W5 are consistent with those reported elsewhere in the literature and show no evidence of spatial variation over orders of magnitude. Moreover, variations in OH abundance measured in the literature occur at smaller length scales than our observations resolve. For example, the variation in OH abundance found by \cite{xu16} near the edge of the Taurus molecular cloud occurs over a total of 2 pc, with each observation covering a 0.12 pc diameter. Our GBT survey has a resolution corresponding to 4.4 pc. The overall average OH abundance over that larger scale is what is relevant to our work.

Calculations of N(H$_2$) determined from \twCO(1-0) data require use of the X-factor. CO column densities are not directly calculated as in the case of the OH analysis, and instead the CO profile integrals are multiplied by the X-factor to estimate the H$_2$ column densities. A thorough discussion of the CO X-factor is presented in \cite{bwl13}, and we note that finding a single high-precision, widely-applicable value for the X-factor is not realistic, as the X-factor can vary up to an order of magnitude depending on location and conditions. However, \cite{bwl13} recommend using a value of $2 \times 10^{20}$ cm$^{-2}$(K*\kmps)$^{-1}$ in the Galactic disk, plus or minus $30 \%$, and we adopt this value for our analysis, while noting that the true uncertainties in molecular gas content derived from CO signals are larger. 

The observational data as well as the calculated values for the column densities from both tracers are listed in Table 2.


\section{Comparing Column Densities from OH and CO}

Here we compare column densities of molecular gas as determined from our OH survey and from CO(1-0) data from the FCRAO survey \cite[]{hbs98} smoothed to the resolution of the GBT survey. We include only features that were detected in emission for at least one of the main line frequencies in our survey, because we find that absorption lines systematically under-predict column densities as compared to emission lines (see Section \ref{sec:unreliable}).

\begin{figure*}[ht!] 
\epsscale{0.9}
\vspace{0.1in}
\begin{center}
\includegraphics[width=0.9\textwidth,angle=0]{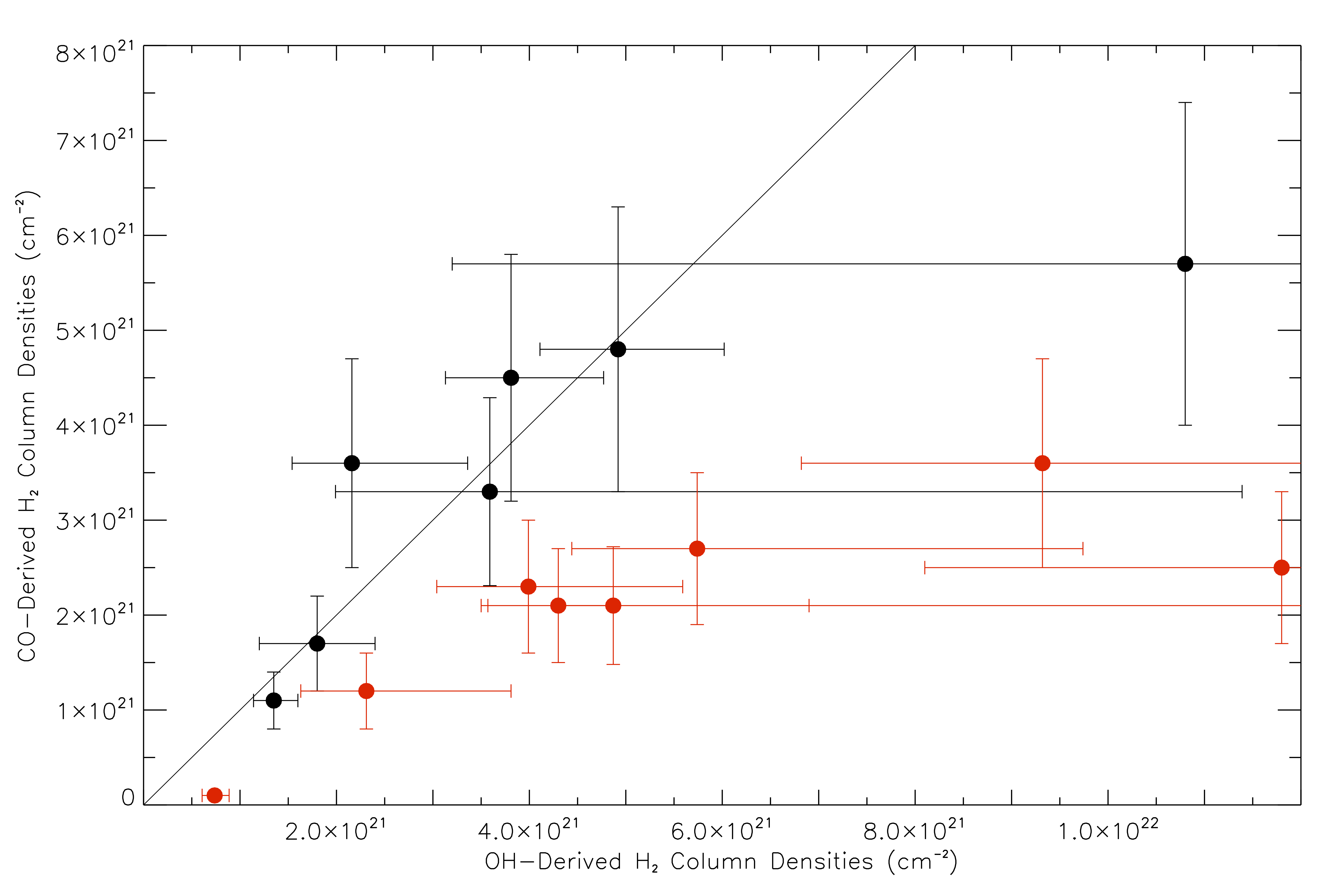} \hspace{0.2in} 
\caption{Derived column densities of N(H$_2$) in W5 for positions within the GBT survey containing OH emission lines at either 1665 MHz or at both main lines using our OH survey data (x axis) and the smoothed FCRAO CO(1-0) data \cite[]{hbs98} (y axis) as tracers. The superimposed diagonal line represents the y = x relation. In 8 out of 15 cases, the OH traces a larger amount of molecular gas than does the CO; these detections are plotted in red. The CO-derived column densities assume the \cite{bwl13} value for the Galactic disk. The OH-derived column densities assume the OH abundance reported by \cite{wes10}. A discussion of assumptions and uncertainties is provided in Sections \ref{sec:Unknowns} and \ref{sec:error}. \label{fig:LongErrorsColDens}}
\end{center}
\end{figure*}

We plot the data from those positions with N(H$_2$) as determined from OH on the x axis and N(H$_2$) as determined from CO(1-0) \cite[]{hbs98} on the y axis. For 8 out of 15 cases, the OH-derived column densities are greater than the CO-derived column densities within the uncertainties. The OH-derived column densities are greater by a factor up to 7. These data are plotted in red in Figure \ref{fig:LongErrorsColDens}. The two molecules trace the same column density within the uncertainties in the other 7 cases, which are plotted in black in Figure \ref{fig:LongErrorsColDens}. In no cases containing OH emission detections are the CO-derived column densities greater than the OH-derived column densities within the uncertainty. Note that the results plotted in Figure \ref{fig:LongErrorsColDens} assume a particular uncertainty in the OH to H$_2$ conversion factor and make use of a standard CO X-factor; true error estimates could be larger than what we assume.

\section{Morphology of Molecular Gas from OH and CO}
\label{sec:morphology}

\begin{figure*}[ht!] 
\epsscale{0.9}
\vspace{0.1in}
\begin{center}
\includegraphics[width=0.8\textwidth,angle=0]{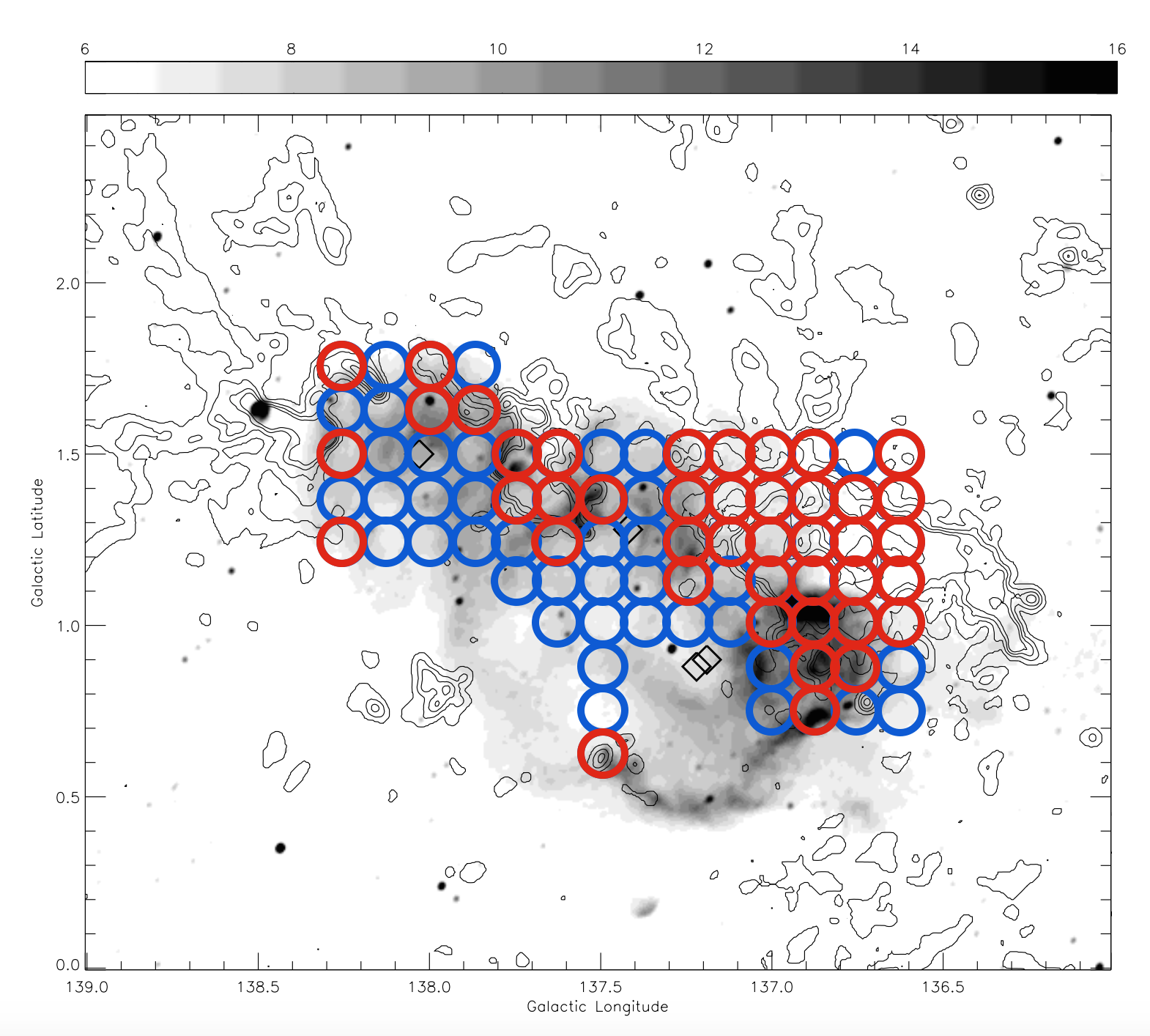} \hspace{0.2in} 
\caption{Positions of our our OH detections in the LSR radial veloccity range -49 to -31 \kmps in red, and positions of observations without detections in blue, overlaid on the map of the W5 region taken directly from \citet[][Figure 2]{km03}. The gray scale indicates the 1420 MHz continuum from the CGPS survey \cite[]{tgp03}, the contours indicate \twCO(1-0) emission from the FCRAO survey \cite[]{hbs98} integrated within the LSR radial velocity range of -49 to -31 km/s, and the 4 diamonds indicate the locations of known O stars in W5 as plotted by \cite{km03}. The circle size representing our observations indicates the $\approx 7\farcm6$ angular resolution FWHM of the GBT. \label{fig:OHMorphologyMore}} 
\end{center}
\end{figure*}

Positions in the survey containing OH detections in W5 within the radial velocity range are plotted in red on Figure \ref{fig:OHMorphologyMore}, which also shows CO(1-0) detections in contours taken directly from \cite{km03}. The morphology of molecular gas near W5 as traced by OH appears to be similar to that traced by CO, even though the column densities derived from the OH detections are in many cases greater than their CO-derived counterparts. All of 44 positions containing OH detections contain CO(1-0) signals, and all but two out of 44 positions containing CO contain OH detections in our survey. The remaining two cases, near the northeast corner of the survey, have $T_C$ near but slightly above $T_{ex}$, which are equivalent within the uncertainties. As such, they are not true non-detections. These positions most likely contain OH that is not easily detectable due to a switch between emission and absorption within the GBT field of view (see Section \ref{sec:unreliable} for an explanation), while the corresponding CO signals are relatively faint and negligible compared to the total gas content of W5. 

The CO and OH data both display two major clouds containing molecular gas in our W5 survey region. These clouds do not conform to the classification of an east and west part of the HII region in W5 discussed by \cite{km03} into W5-E and W5-W. Rather, the molecular gas according to both tracers contains a large cloud on the west side of W5 that extends beyond the W5 continuum to the north and west, and a smaller cloud closer to the eastern end of the W5 continuum, located near the boundary between W5-E and W5-W. The edge of another cloud extending outside of W5 exists at the easternmost boundary of the survey. There are voids containing no molecular gas detections near the centers of W5-E and W5-W, most likely resulting from ionization and wind produced by the O and B stars in the centers of those regions (see Figure \ref{fig:OHMorphologyMore}).

\section{Mass Estimates of Molecular Gas from OH and CO}
\label{sec:mass}

We obtain mass estimates of molecular gas in the survey region from our column density data. However, more than half of the OH detections manifest as absorption lines, which are not suitable for inclusion in the mass estimate because they systematically under-predict column densities (see Section \ref{sec:unreliable}). Thus we only perform mass estimates where there is a usable OH emission detection at at least one main line frequency. Fifteen out of 42 pointings ($\sim 36 \%$) with OH detections meet this criterion and are included in the mass estimate. 

To estimate mass, we find the column density in each pixel in the survey containing an OH emission detection at either of the main line frequencies, and calculate the mass that would fit within each pixel if the majority of molecular gas particles are H$_2$ molecules at the distance to W5 of 2 kpc. Then we sum all of those pixels to find the mass of molecular gas in the portion of W5 being studied.

For the subset of the positions that contain OH emission detections in some form, we calculate a total mass of $1.7 \times 10^4 M_{\odot}$ + $0.6 \times 10^4 M_{\odot}$ or - $0.2 \times 10^4 M_{\odot}$. The corresponding CO data gives a total mass of $9.9 \times 10^3 M_{\odot} \pm 0.7 \times 10^3 M_{\odot}$.  


Since fewer than half of the pointings containing OH detections at W5 radial velocities are able to be used in the mass estimate, the total mass in W5 could be roughly two to three times the mass that we calculated from column densities derived from OH emission lines.

\section{Calculations and Uncertainties}
\label{sec:Calculations and Uncertainties}
\subsection{Unknowns Involved}
\label{sec:Unknowns}

Here we consider the details, uncertainties, and assumptions in calculating N(OH) and N(H$_2$) for the simpler case of emission. First, we must assume a value of $T_C$ in the background of any OH gas when calculating column densities from observations. The line of sight position of the OH feature and any uncertainty in that position are relevant because $T_C$ depends on the position of the gas relative to the source of the continuum.


Second, the signal being measured is in fact the weighted integral of $\int{T_b(\nu)d\nu}$ over the GBT point spread function. Therefore, any high-resolution spatial variation of the signal within the telescope field of view cannot be detected and is an unknown quantity. If the $T_{ex}$ and $T_{C}$ values were constant over the telescope beam, there would be no problem, but $T_C$ at least certainly varies spatially at high resolution compared to the GBT point spread function. While $T_{ex}$ in principle could vary as well, our results in \cite{ea18} show no evidence of variation in $T_{ex}$ at scales comparable to the GBT point spread function above the uncertainty. Variation in $T_C$ poses a considerable problem, because N(OH) as well as $T_C$ can both vary within the GBT beam. The resulting spectral signal, which is the integral of $\int{T_b(\nu)d\nu}$ over the point spread function, depends on how the distributions of N(OH) and $T_C$ overlap. Methods for addressing these unknowns and mitigating them when possible are discussed in Sections \ref{sec:Line of Sight} and \ref{sec:Filling Factor}.

Lastly, we must point out that we are assuming the \cite{wes10} value for the OH to H$_2$ ratio in W5 despite the fact that this value was not measured in W5. Although we have demonstrated in Section \ref{sec:total} that it is a reasonable value, it is important to remember that this value is nevertheless an assumption.

\subsection{Line of Sight Geometry}
\label{sec:Line of Sight}

Much of our analysis of the OH line profiles and calculations of excitation temperatures and column densities assumes that the OH gas is located in the foreground of the W5 continuum source. Naturally, not all of the OH in the vicinity of W5 will be located in the foreground. A good starting point is to assume that most of the OH will be either in the foreground or the background of the W5 HII region, because OH gas located in the midst of the HII region would be photo-dissociated and ionized \cite[e.g.][]{abj09}. This may not in fact be entirely true, as we shall see, because a clumpy and irregularly shaped HII region may allow for pockets of molecular gas to exist partway through the HII region continuum source along the line of sight. 

For the majority of positions in the survey at which OH was detected, it appears that there is OH in the foreground. This finding is clear for examples at which OH absorption was detected at either main line frequency since we know that $T_{ex}$ is not low enough to allow for OH absorption when the W5 continuum is not in the background. There are also positions detected in emission at both main line frequencies at which evidence implies that the OH is located in the foreground. This evidence comes from examining the line ratios, the $T_C$ values, and the $T_{ex}$ values. Main line emission profile integral ratios departing from 5:9 could be a result of the fact that $T^{65}_{ex} > T^{67}_{ex}$. As described in \cite{ea18}, our observed gradient of line ratio departures from 5:9 over the northwest part of the survey are consistent with a model for which the OH is in the foreground.

At several positions, there is evidence of OH in both the foreground and the background of the W5 continuum. This type of spectrum manifests as an emission line and an absorption line next to each other or even overlapping slightly, creating a complicated spectrum. Radial velocities and widths of the lines tend to vary enough within different parts of the survey that a perfect overlap, which would be difficult to separate, is unlikely. In cases containing OH in the foreground as well as the background, we choose to include only the foreground features in our study to simplify the analysis, as $T_C$ and $T_{ex}$ values are known more accurately. 

That accounts for all of the OH features used in our analysis, but there are two remaining cases evidenced in the survey. At a few positions in the eastern portion of the survey, we find that the OH is detected in emission at both main line frequencies at the 5:9 ratio, despite the fact that $T_C$ at that position is is above the excitation temperature for both main lines. If the gas were in the foreground of the continuum, both lines would have been detected in absorption. Instead, we suppose that the gas is located entirely in the background, so that the value of $T_C$ behind the gas is much lower.

There are three adjacent positions in the survey with evidence of OH located somewhere in the midst of the continuum along the line of sight. These positions contain absorption detections at 1667 MHz and no OH detection at 1665 MHz. This finding would imply that $T_C$ behind the OH at those positions is equal to the $T_{ex}$ at 1665 MHz. However, the $T_C$ values at those positions are greater than that value. Our proposed explanation is that these are cases for which OH is located in a shielded pocket within the HII region producing the W5 continuum, neither in the foreground nor in the background. It is unlikely that $T_{ex}$ variation could explain these spectra because those three positions are adjacent and yet would have $T^{65}_{ex}$ values varying significantly more than $T_{ex}$ was found to vary in \cite{ea18}. 


\subsection{Effects of Structure Inside GBT Beam for Emission Lines}
\label{sec:Filling Factor}


In Section \ref{sec:Unknowns}, we described how the observed emission spectrum at a given position depends not simply on the beam-averaged values of N(OH) and $T_C$, but on the high resolution spatial distributions of N(OH) and $T_C$ within the GBT field of view, and how those distributions overlap.


Thus, the relevant value to insert in Equation \ref{eq:emission} is not the beam-averaged value of $T_C$, but rather the average value of $T_C$ convolved with distribution of N(OH) within the GBT beam, which we call $T^{optimal}_C$. 

\begin{figure*}[ht!] 
\epsscale{0.9}
\vspace{0.1in}
\begin{center}
\includegraphics[width=0.65\textwidth,angle=0]{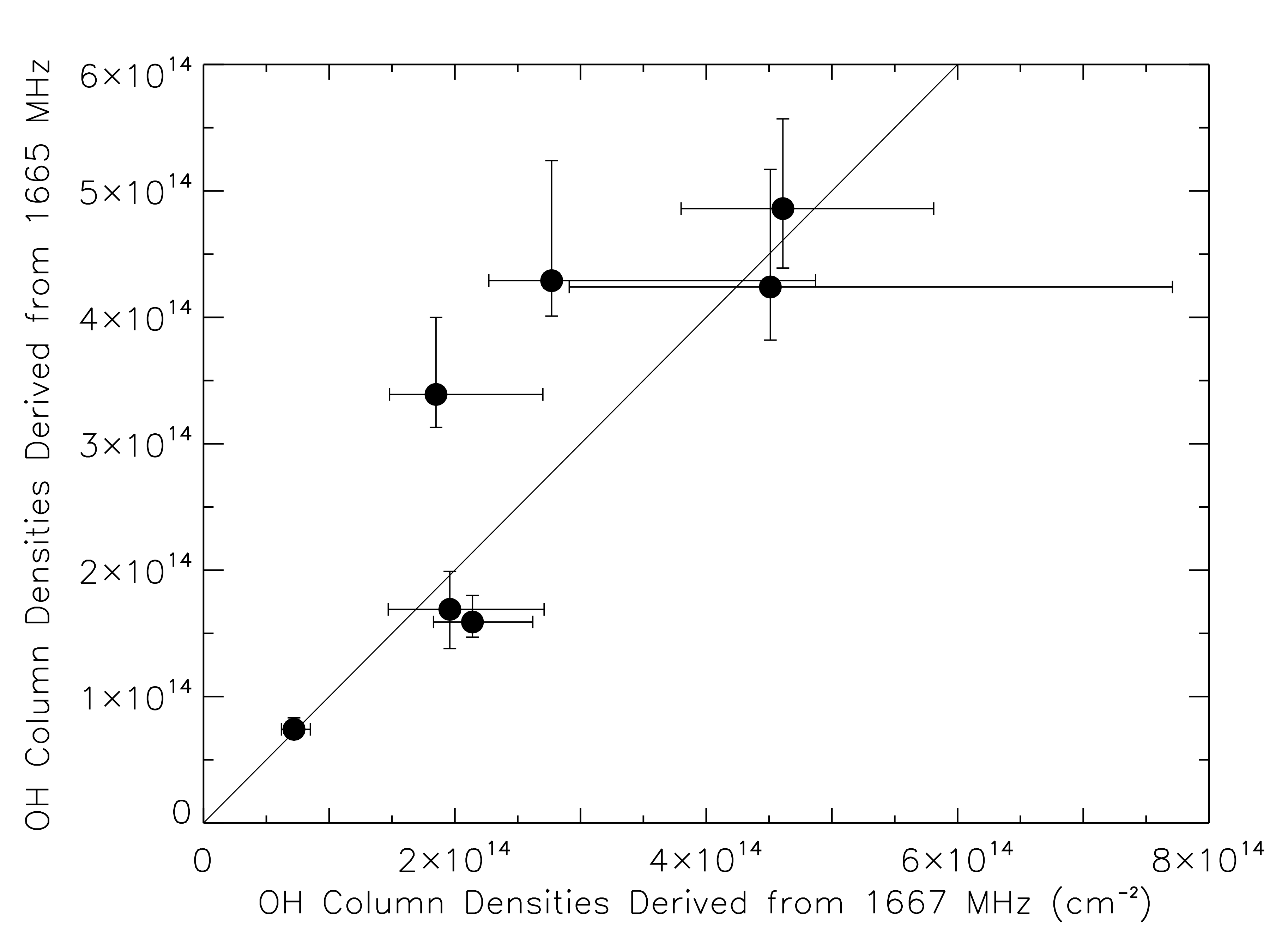} \hspace{0.2in} 
\caption{Column densities of OH as calculated from the 1667 MHz OH emission profiles compared to the OH column densities for the same positions as calculated from the 1665 MHz OH emission profiles for all observed cases in which both 1665 MHz and 1667 MHz contain emission profiles within the W5 radial velocity range. A diagonal y = x line is superimposed to indicate where the column densities as calculated from 1665 MHz and from 1667 MHz emission lines are the same. The error bars contain uncertainties in the profile integrals, continuum temperature, and the excitation temperature values from \cite{ea18}. The plot demonstrates that the 1665 MHz and 1667 MHz emission lines yield generally consistent column density results when the two different values for the main line excitation temperatures are taken into account. \label{fig:6567}}
\end{center}
\end{figure*}

We propose a method to determine $T^{optimal}_C$ for a subset of observations. For the rest, we will estimate the uncertainty posed by using the beam-averaged $T_C$ instead of the unknown $T^{optimal}_C$.  
There are several positions in our survey at which OH emission was detected at both main line frequencies. These cases allow us to compare the column densities calculated independently from the two main lines at the same position in the sky. If the $T_{ex}$ and $T_C$ input parameters are correct, then the two main lines at the same position should yield equal column density results, within the statistical uncertainty. However, in general, there is some scatter in the column density comparison when the measured $T_{ex}$ values (see \cite{ea18}; Appendix \ref{app:Tex}) and the beam-averaged $T_C$ are used as input parameters, as seen in Figure \ref{fig:6567}. 


To find $T^{optimal}_C$ for these positions, we begin by assuming the $T_{ex}$ values as given. Then we can write Equation \ref{eq:emission} for both 1665 MHz and for 1667 MHz at the same position, and set the equations equal to each other, with the goal of solving for $T^{optimal}_C$:
\begin{displaymath}
N(OH) = C_{65}\frac{T^{65}_{ex}}{T^{65}_{ex} - T^{optimal}_C}\int{T^{65}_b(\nu)d\nu} 
\end{displaymath}
\begin{equation}
= C_{67}\frac{T^{67}_{ex}}{T^{67}_{ex} - T^{optimal}_C}\int{T^{67}_b(\nu)d\nu}.
\label{eq:setequal}
\end{equation}

A similar line of reasoning leads to a method to constrain the excitation temperature values, as described in Appendix \ref{app:Tex}. These values are $T^{65}_{ex} = 5.87 + 0.43$ or - $0.37$ K, and $T^{67}_{ex} = 5.13 + 0.17$ or - $0.22$ K.

For the emission detections at 1665 MHz that do not contain corresponding emission at 1667 MHz, it is not possible to determine $T^{optimal}_C$, so we assume the beam-averaged $T_C$. To estimate this contribution to the uncertainty in N(OH), we implement Equation \ref{eq:emission} with the minimum and maximum values of $T_C$ within the GBT field of view at the CGPS resolution, scaling for 1-$\sigma$ uncertainties. 



\subsection{Summary of Error Sources}
\label{sec:error}

Here we summarize the four error sources in N(H$_2$) calculated from OH emission observations. First, there is the statistical noise uncertainty in the profile integral measurements from the GBT. Second, there is uncertainty from $T_{ex}$ (see Appendix \ref{app:Tex}, Section \ref{sec:Filling Factor}). Third, there is uncertainty in $T_C$, which can come in two forms. For observations containing emission at both main line frequencies, an improved accuracy $T^{optimal}_{C}$, resulting from variation within the GBT beam of both N(OH) and $T_C$ distributions, can be calculated as described in Section \ref{sec:Filling Factor}. The uncertainty in $T^{optimal}_C$ depends on the precision to which $T^{optimal}_C$ could be determined as constrained by uncertainty in the profile integrals and $T_{ex}$. For spectra containing emission only at 1665 MHz, $T^{optimal}_C$ cannot be calculated, and the resulting uncertainty in $T_C$ is discussed in Section \ref{sec:Filling Factor}. Fourth, there is uncertainty in the conversion from N(OH) to N(H$_2$) provided by \cite{wes10}. 

For corresponding CO-derived N(H$_2$) values, there are two sources of uncertainty: the statistical noise uncertainty in the CO profile integrals from the smoothed FCRAO survey data \cite[]{hbs98}, and the uncertainty in the value of the X-factor for gas within the Galactic disk, which is provided by \cite{bwl13}.

\section{Discrepancies with Absorption Line Column Densities}
\label{sec:unreliable}



We have not used spectra detected in absorption in our column density calculations. Here, we examine a subset of spectra that contain absorption at 1667 MHz and emission at 1665 MHz, and compare N(OH) as calculated from those two lines at each position. There are five such examples in the survey data. 

We apply Equation \ref{eq:emission} for the 1665 MHz emission line and Equation \ref{eq:absorption} for the 1667 MHz absorption line, with the appropriate coefficients for the respective line frequencies. The result is that the absorption line yields a column density one to two orders of magnitude lower than does the emission line. The absorption-derived column densities appear more likely to be the inaccurate ones, because the emission-derived column densities are mostly at the same order of magnitude as corresponding CO-derived column densities even though they are sometimes greater by a factor between 1 and 10. On the other hand, the absorption-derived column densities are always at least one order of magnitude below the CO-derived equivalents and sometimes more. Apparently, the absorption lines in our survey systematically under-predict the column densities by a factor of 10 to 100, as seen in Table \ref{table:AbsorptionDiscrepancy}. 

\begin{table*}[ht!]
\begin{center}

\caption{OH Column densities as calculated from absorption at 1667 MHz and emission at 1665 MHz at the same coordinates \label{table:AbsorptionDiscrepancy}}
\begin{tabular}{l l c c c}
\hline
 l & b & 1667 Absorption N(OH) (cm$^{-2}$) & 1665 Emission N(OH) (cm$^{-2}$) & Ratio\\
\hline
136.875 & 1.25 & 8.5 $\times$ 10$^{12}$ & 2.3 $\times$ 10$^{14}$ & 0.037 \\
137.0 & 1.25 & 1.3 $\times$ 10$^{13}$ & 1.1 $\times$ 10$^{15}$ & 0.011 \\ 
137.0 & 1.375 & 1.2 $\times$ 10$^{13}$ & 9.8 $\times$ 10$^{14}$ & 0.012 \\
137.125 & 1.25 & 8.5 $\times$ 10$^{12}$ & 9.7 $\times$ 10$^{14}$ & 0.0089 \\
\end{tabular}
\end{center}
\end{table*}

We propose that absorption lines systematically under-predict the column density as compared to emission lines because two conditions must be met for an absorption line to occur: there must be OH present to absorb, and there must be a sufficiently strong background continuum present to be absorbed. Only where these two conditions overlap does absorption occur, and only the fraction of OH that overlaps with sufficiently strong continuum temperatures will be detected. 

We have already established in Sections \ref{sec:Unknowns} and \ref{sec:Filling Factor} that OH probably covers just a fraction of the GBT field of view. There is much evidence in the literature to support this claim. HI absorption interferometry observations have provided evidence of small-scale structure and variation in opacity from parsec scale down to a few AU \cite[]{brogan05, d00, fg01, frail94, goss08, roy10, roy12}. Comparison to single-dish observations, moreover, suggests that the observed variation in opacities cannot result from variation in spin temperature and more likely results from structural variation \cite[]{brogan05}. There is also evidence of small-scale variational structure and clumpiness in molecular gas from absorption interferometry observations of H$_2$CO found by \cite{goss84, rg02} in front of Cas A, and \cite{mm95, marscher93} in front of 3C111, BL Lac, and NRAO 150. In OH absorption, \cite{bc86} found similar structure in front of Cas A. These studies all indicate the likelihood that the OH filling factor is low; filling factors of about 10 percent have been found at small scales for HI \cite[]{brogan05} and could plausibly be similar or even smaller for OH. The small scale variation and clumping of OH, combined with higher resolution structure of the background continuum which can be viewed in the CGPS \cite[]{tgp03} 1420 MHz continuum maps, suggest that absorption will only occur for a fraction of the OH within a GBT field of view even if the beam averaged $T_C$ is greater than $T_{ex}$. The result would be that absorption lines will systematically underestimate N(OH), just as we find in our data.

Additionally, OH from portions of the GBT field of view containing lower $T_C$ values may exhibit OH emission profiles, despite the beam-averaged $T_C$ being greater than $T_{ex}$. These emission components may not dominate the beam-averaged spectrum, but can diminish the strength of the absorption profile in the beam-averaged spectrum.


Thus there are two main reasons why absorption profiles will under-predict the column densities of OH: they only account for a fraction of the OH that is present, and the remaining OH may further reduce the signal strength by adding emission profiles on top of it. 

\begin{figure*}[ht!] 
\epsscale{0.9}
\vspace{0.1in}
\begin{center}
\includegraphics[width=0.3\textwidth,angle=0]{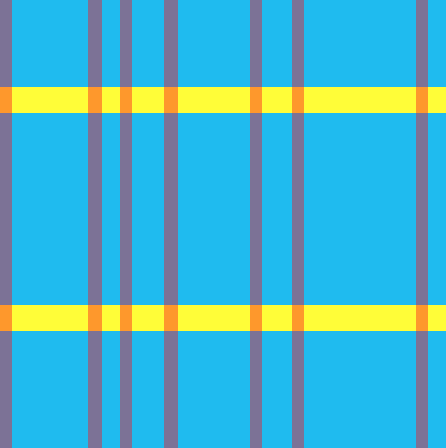} \hspace{0.2in} 
\caption{Diagram of the type of model used to simulate the effects of high resolution spatial structure of OH and $T_C$ on resulting beam-averaged spectra. The blue regions represent relatively low values of $T_C$ and N(OH), red regions represent higher values of $T_C$, and yellow regions represent higher values of N(OH). Absorption profiles occur at pixels containing an overlap of red and yellow regions, which only includes a fraction of the total OH content in the model. Moreover, pixels in the yellow regions containing high OH content but lower $T_C$ values produce emission profiles, which contribute to the beam-averaged spectrum along with the absorption profiles from the overlapping red and yellow regions. The net effect is a beam-averaged absorption profile that predicts a lower value of beam-averaged N(OH) than is truly present in the model. This diagram does not necessarily display the ratios of blue, red, and yellow regions that produced results comparable to our measurements, but is simply intended to illustrate the model generally.\label{fig:SquareSimulation}}
\end{center}
\end{figure*}

In order to test the plausibility of these arguments, we create a simulation containing filaments of OH and independent variation of elevated continuum temperature over a square telescope beam, and calculate what the detected line profile would be at each pixel within the square. We calculate the total beam-averaged predicted line profile by averaging the results from all of the pixels, and we calculate the beam-averaged $T_C$ as well. Next, we compare the column density calculated from the beam-averaged predicted spectrum to the input column density value used in the simulation. We perform this simulation several times while varying the input parameters for the model. For a given beam-averaged column density of OH, the predicted detection results in lower OH column densities, which can be one to two orders of magnitude less for certain input parameter values, comparable to the observed results. Although the true telescope beam is not square, the difference in the model would be negligible. We conclude that our explanation for the reduced values of OH column densities calculated from absorption lines is plausible. A diagram of the model is shown in Figure \ref{fig:SquareSimulation}.

For the two positions in the northeast corner of the survey where faint CO is detected without a corresponding OH detection, beam-averaged $T_C$ values are just slightly above $T_{ex}$ and $T_C$ varies above and below $T_{ex}$ within the field of view. At these two positions, we suggest that the OH absorption spectrum is so diminished in this way as to be undetectable above the noise level.

Future interferometry observations of OH 18-cm absorption at these positions could provide the data needed to accurately calculate column densities. Currently, we do not have this data, and owing to differences in structure at each position, there does not appear to be a way to correct for the column density discrepancies without introducing overwhelming uncertainties.

However, this exercise does allow us to estimate the size of the OH filling factors for these observations. The filling factor is probably roughly a few to 10 percent and varies between the observations.

We note that just as it is possible for absorption profiles to be mixed with some emission profiles when $T_C \approx T_{ex}$, it is also possible for emission profiles to be mixed with some absorption profiles when $T_C \approx T_{ex}$. Therefore, a few of the column densities calculated from emission lines may be lower limits on the true column density values.

\section{Discussion}
\label{sec:Discussion}


The results presented in this paper as well as past studies \cite[e.g.][]{ahe15, nguyen18} imply that OH is a viable tracer for interstellar molecular gas. Although our results suggest that OH might trace more molecular gas in W5 than does CO, additional uncertainties regarding the CO X-factor and the OH abundance prevent the results from providing conclusive evidence for the presence of CO-dark molecular gas in W5. In any case, no position in the W5 survey containing OH is entirely CO-dark. The nearly one-to-one correspondence between OH and CO distributions in the W5 survey differs from the results found by \cite{ahe15} for a quiescent region centered on $l$ = $105.0^{\circ}$, $b$ = $1^{\circ}$, where fewer than half of positions with OH detections contained CO counterparts detected in the \cite{dht01} CO data. Thus we do find conclusive evidence of CO-dark molecular gas in that quiescent region. One possible explanation for this difference could be as follows. The CO(1-0) signal only appears when the gas exceeds a critical volume density. Turbulence in the W5 star-forming region could lead to the formation of dense concentrations in portions of the molecular gas clouds, which may not be common in the quiescent region. 





\section{Conclusions}

Our goal was to study the molecular gas content in a star-forming region using OH 18-cm lines as a tracer, and to compare it to the view as traced by CO(1-0). We calculated column densities of molecular gas in W5 using a GBT OH 18-cm main line grid survey of W5. From this survey, we conclude that OH traces a similar morphology of molecular gas as does CO near W5. At positions where at least one main line frequency is observed in emission, OH traces $1.7 \times 10^4 M_{\odot}$ (+ $0.6 \times 10^4 M_{\odot}$ or - $0.2 \times 10^4 M_{\odot})$ of molecular gas, whereas CO traces $9.9 \times 10^3 M_{\odot} \pm 0.7 \times 10^3$. 

During the course of the project, we also determined that the OH absorption lines in our survey systematically underestimate the column densities by 1 to 2 orders of magnitude. We are able to explain the this discrepancy by considering the extent to which the OH filling factor overlaps with areas of elevated continuum temperature, and modeling suggests an OH filling factor of a few to 10 percent for the survey region. Future interferometry observations could remove this discrepancy by providing higher spatial resolution OH absorption spectra. Filling factor considerations also allow a method to constrain excitation temperature values to $T^{65}_{ex} = 5.87 + 0.43$ or - $0.37$ K and $T^{67}_{ex} = 5.13 + 0.17$ or - $0.23$ K. The total molecular gas content of W5 is probably about two to three times our reported mass estimates because only $36 \%$ of the OH detections contained emission lines and were used in the mass calculations.







\acknowledgements

We thank W. Miller Goss for discussions and suggestions about this project and this paper, which led to improvements in its clarity. We thank David E. Hogg, a collaborator and coauthor on much of our research program, for his contributions to the overall project as well as many discussions, especially regarding the details of GBT observing and data reduction. We thank the anonymous referee for comments that led to revisions to improve the paper. We are grateful to the staff at the Green Bank Observatory for their advice and assistance with the operation of the GBT, and for the development and support of the GBTIDL data analysis software. The National Radio Astronomy Observatory is a facility of the National Science Foundation operated under cooperative agreement by Associated Universities, Inc. The Green Bank Observatory is a facility of the National Science Foundation operated under cooperative agreement by Associated Universities, Inc. The work reported here has been partially supported by the Director's Research Funds at the Space Telescope Science Institute. Support for this work was also provided by the NSF through the Grote Reber Fellowship  Program administered by Associated Universities, Inc./National Radio Astronomy Observatory. The research presented in this paper has used data from the Canadian Galactic Plane Survey, a Canadian project with international partners, supported by the Natural Sciences and Engineering Research Council of Canada. This research has benefited from the SAOImage software provided by the Harvard-Smithsonian Center for Astrophysics; we are grateful to Bill Joye for his advice on smoothing data using SAOImage.


\appendix

\section{Further Constraints on Excitation Temperatures} \label{app:Tex}

The filling factor analysis in Section \ref{sec:Filling Factor} assumed the excitation temperature values as input parameters, but it also provides a way to further constrain their values. In \cite{ea18}, we measured the excitation temperatures to be $T_{ex}^{65} = 6.0 \pm 0.5$ K, $T_{ex}^{67} = 5.1 \pm 0.2$ K, and $T_{ex}^{65} - T_{ex}^{67} = 0.9 \pm 0.5$ K. If instead of using the \cite{ea18} excitation temperature values as inputs, we were to choose an arbitrary pair of values for $T^{65}_{ex}$ and $T^{67}_{ex}$, then a different value of $T_C$ would result. However, not any arbitrary value of $T_C$ is physically possible. There is a range of $T_C$ values that exists within the field of view of a given GBT observation, and only solutions for $T_C$ that exist between the minimum and maximum value of $T_C$ within the GBT field of view are possible solutions. Therefore, any combination of input $T_{ex}$ values that leads to a solution for $T_C$ outside of the extrema for that observation can be ruled out. We can use this principle to set up a system of inequalities, one for the maximum $T_C$ and one for the minimum $T_C$ for each observation containing OH emission at both main line frequencies. The solution of the system of inequalities is the phase space of possible values of $T^{65}_{ex}$ and $T^{67}_{ex}$. Since the system of inequalities involved five observations covering an angular region of W5 greater than a single GBT point spread function, any potential low level variation in $T_{ex}$ at that scale is still incorporated within the uncertainties on the new constrained values. We have therefore constrained the values of $T_{ex}$ in W5 to 

$T^{65}_{ex} = 5.87$ (+ 0.43 or - 0.37) K, 

$T^{67}_{ex}$ = 5.13 (+ 0.17 or - 0.23) K, and 

$T^{65}_{ex} - T^{67}_{ex}$ = 0.74 + (0.26 or - 0.22) K. 

These values are an improvement in precision over the \cite{ea18} values.

\begin{center}
\begin{longrotatetable}
\begin{deluxetable*}{l l c c c c c c c}

\tablecaption{Complete table of data from the GBT survey of W5. Listed for each position in the survey are the galactic longitude, galactic latitude, 1665 MHz profile integral, 1667 MHz profile integral, continuum temperature, calculated column density of OH, \twCO profile integral from \cite{hbs98}, calculated molecular gas column density from OH, and calculated molecular gas column density from CO. In the OH 1665 and 1667 MHz profile integral columns, if the feature was detected in absorption rather than emission, then instead of a profile integral, the integral of the optical depth is listed preceded by "$\tau = $." In cases for which only absorption features were detected at both OH main line frequencies, we provide parenthesis around the calculated column density values because these values are systematically underestimated. If no detection was made above the noise level of the spectra, we list upper limits for column densities at that position. If a W5 feature was determined to be somewhere other than the foreground, we indicate this in the table because then the listed value of $T_C$ may not be accurate for that feature.}
\tablewidth{600pt}
\label{table:surveyresults}
\tabletypesize{\tiny}
\tablehead{
\colhead{l} & \colhead{b} & \colhead{1665 MHz (K*km/s or $\tau$)} & \colhead{1667 MHz (K*km/s or $\tau$)} & \colhead{$T_C$ (K)} & \colhead{N(OH) (cm$^{-2}$)} & \colhead{\twCO\ Profile Integral (K*\kmps)} & \colhead{N(\Htwo)$_{OH}$ (cm$^{-2}$)} & \colhead{N(\Htwo)$_{CO}$ (cm$^{-2}$)} 
}
\startdata
$136.625^{\circ}$ & $0.750^{\circ}$ & ... & ... &$5.4$ + $0.4$ & no upper limit & ... & no upper limit &$< 2 \times 10^{20}$ \\
 & & & &or - $0.2$ & & & & \\
$136.625^{\circ}$ & $0.875^{\circ}$ & ... & ... &$6.0$ + $0.9$ &$(< 4 \times 10^{12})$ & ... &$(< 4 \times 10^{19})$ &$< 2 \times 10^{20}$ \\
 & & & &or - $0.4$ & & & & \\
$136.625^{\circ}$ & $1.000^{\circ}$ & $\tau = 0.002 \pm 0.002$ & $\tau = 0.007 \pm 0.002$ & $5.5$ + $0.6$ & ($6.6 \times 10^{12}$) &  $6.2 \pm 0.1$ &$(6.2 \times 10^{19})$ & $1.2 \times 10^{21} \pm 4 \times 10^{20}$ \\ 
 & & & &or - $0.2$ & & & & \\ 
$136.625^{\circ}$ & $1.125^{\circ}$ & $0.239 \pm 0.003$ & $0.241 \pm 0.003$ & $4.42 \pm 0.1$& $4.01 \times 10^{14} + 8.5 \times 10^{13}$ & $22.2 \pm 0.1$ & $3.81 \times 10^{21}$ + $9.6 \times 10^{20}$ & $4.5 \times 10^{21} \pm 1.3 \times 10^{21}$ \\ 
& & & & &or - $4.6 \times 10^{13}$ & &or - $6.8 \times 10^{20}$ & \\ 
$136.625^{\circ}$ &$1.250^{\circ}$ &$0.334 \pm 0.003$ &$0.363 \pm 0.003$ & $4.30 \pm 0.1$ &$5.17 \times 10^{14} + 8.5 \times 10^{13}$ &$24.2 \pm 0.1$ & $4.92 \times 10^{21} + 1.1 \times 10^{21}$ &$4.8 \times 10^{21} \pm 1.5 \times 10^{21}$ \\ 
& & & & &or - $4.7 \times 10^{13}$ & &or - $8.1 \times 10^{20}$ & \\ 
$136.625^{\circ}$ &$1.375^{\circ}$ &$0.116 \pm 0.003$ &$0.187 \pm 0.003$ &$4.1 \pm 0.09$ &$1.9 \times 10^{14} \pm 5 \times 10^{13}$ &$8.3 \pm 0.2$ &$1.8 \times 10^{21} \pm 6 \times 10^{20}$ &$1.7 \times 10^{21} \pm 5 \times 10^{20}$ \\ 
& & & & & & & & \\ 
$136.625^{\circ}$ &$1.500^{\circ}$ &$0.057 \pm 0.003$ &$0.069 \pm 0.002$ &$4.08 \pm 0.1$ &$7.75 \times 10^{13} + 1.1 \times 10^{13}$ &$0.5 \pm 0.1$ &$7.37 \times 10^{20} + 1.5 \times 10^{20}$ &$1.0 \times 10^{20} \pm 3.6 \times 10^{19}$ \\ 
& & & & & or - $9.0 \times 10^{12}$& &or - $1.3 \times 10^{20}$& \\ 
$136.750^{\circ}$ &$0.750^{\circ}$ & ... & ... &$7.1$ + $3.3$&$< 4 \times 10^{14}$ & ... &$< 4 \times 10^{21}$ &$< 2 \times 10^{20}$ \\
 & & & &or - $0.7$ & & & & \\
$136.750^{\circ}$ &$0.875^{\circ}$ &$\tau = 0.015 \pm 0.002$ &$\tau = 0.017 \pm 0.002$ &$8.4$ + $0.7$ &$(2.8 \times 10^{13})$ &$4.8 \pm 0.2$ &$(2.7 \times 10^{20})$ &$9.6 \times 10^{20} \pm 2.9 \times 10^{20}$ \\ 
& & & &or - $0.5$ & & & & \\ 
$136.750^{\circ}$ &$1.000^{\circ}$ &$\tau = 0.037 \pm 0.002$ &$\tau = 0.059 \pm 0.002$ &$8.0 \pm 1.0$ &$(8.0 \times 10^{13})$ &$16.4 \pm 0.1$ &$(7.6 \times 10^{20})$ &$3.3 \times 10^{21} \pm 9.8 \times 10^{20}$ \\ 
& & & & & & & & \\ 
$136.750^{\circ}$ &$1.125^{\circ}$ &$0.073 \pm 0.003$ & --&$5.4 + 1.0$ &$3.77 \times 10^{14} + 8.0 \times 10^{14}$ &$16.4 \pm 0.2$ &$3.59 \times 10^{21} + 7.8 \times 10^{21}$&$3.3 \times 10^{21} \pm 9.9 \times 10^{20}$ \\ 
& & & &or - 0.5 &or - $1.6 \times 10^{14}$ & &or - $1.6 \times 10^{21}$ & \\ 
$136.750^{\circ}$ &$1.250^{\circ}$ &$0.219 \pm 0.003$ &$0.130 \pm 0.003$ &$4.83 \pm 0.1$ &$5.12 \times 10^{14} + 7.7 \times 10^{14}$ &$10.3 \pm 0.2$ &$4.87 \times 10^{21} + 7.3 \times 10^{21}$ &$2.1 \times 10^{21} \pm 6.2 \times 10^{20}$\\ 
& & & & &or - $1.2 \times 10^{14}$ & &or - $1.3 \times 10^{21}$ & \\ 
$136.750^{\circ}$ &$1.375^{\circ}$ &$0.277 \pm 0.003$ &$0.195 \pm 0.003$ &$4.75 \pm 0.1$ &$6.03 \times 10^{14} + 4.1 \times 10^{14}$ &$13.6 \pm 0.2$ &$5.74 \times 10^{21} + 4.0 \times 10^{21}$ &$2.7 \times 10^{21} \pm 8.0 \times 10^{20}$ \\ 
& & & & &or - $1.1 \times 10^{14}$ & &or - $1.3 \times 10^{21}$ & \\ 
$136.750^{\circ}$ &$1.500^{\circ}$ & ... & ... &$4.2$ + $0.2$ & $< 6 \times 10^{12}$ & ... & $< 6 \times 10^{19}$ &$< 2 \times 10^{20}$ \\
 & & & &or - $0.1$ & & & & \\
$136.875^{\circ}$ &$0.750^{\circ}$ &$\tau = 0.004 \pm 0.002$ &$\tau = 0.008 \pm 0.002$ &$8.8$ + $4.6$&$(9.6 \times 10^{12})$ &$1.2 \pm 0.2$ &$(9.1 \times 10^{19})$ &$2.4 \times 10^{20} \pm 8 \times 10^{19}$ \\ 
& & & &or - $1.1$ & & & & \\ 
$136.875^{\circ}$ &$0.875^{\circ}$ &$\tau = 0.023 \pm 0.002$ &$\tau = 0.035 \pm 0.002$ &$9.2$ + $0.7$&$(4.9 \times 10^{13})$ &$9.2 \pm 0.2$ &$(4.6 \times 10^{20})$ &$1.8 \times 10^{21} \pm 6 \times 10^{20}$ \\ 
& & & &or - $0.8$ & & & & \\ 
$136.875^{\circ}$ &$1.000^{\circ}$ &$\tau = 0.054 \pm 0.002$ &$\tau = 0.097 \pm 0.002$&$10.7$ + $1.9$&$(1.2 \times 10^{14})$ &$24.8 \pm 0.2$ &$(1.2 \times 10^{21})$ &$5.0 \times 10^{21} \pm 1.5 \times 10^{21}$ \\ 
& & & &or - $0.7$ & & & & \\ 
$136.875^{\circ}$ &$1.125^{\circ}$ &$\tau = 0.050 \pm 0.002$ &$\tau = 0.083 \pm 0.002$ &$7.9$ + &$(1.1 \times 10^{14})$ &$57.4 \pm 0.2$ &$(1.0 \times 10^{21}$ &$1.1 \times 10^{22} \pm 3 \times 10^{21}$ \\ 
& & & &or - & & & & \\ 
$136.875^{\circ}$ &$1.250^{\circ}$ &$0.052 \pm 0.002 $&$0.072 \pm 0.003$ &$5.1 \pm 0.3$ &$2.27 \times 10^{14} + 1.2 \times 10^{14}$ &$18.1 \pm 0.2$ &$2.16 \times 10^{21} + 1.2 \times 10^{21}$&$3.6 \times 10^{21} \pm 1.1 \times 10^{21}$ \\ 
& &and $\tau = 0.007 \pm 0.002$ & & &or - $5.7 \times 10^{13}$ & &or - $6.2 \times 10^{20}$ & \\ 
$136.875^{\circ}$ &$1.375^{\circ}$&$0.17 \pm 0.003$& ... &$4.9$ + 0.5 &$4.26 \times 10^{14}$ + $2.8 \times 10^{14}$ &$10.6 \pm 0.2$ &$4.3 \times 10^{21}$ + $2.6 \times 10^{21}$ &$2.1 \times 10^{21} \pm 6 \times 10^{20}$ \\ 
& & & &or - 0.2 &or - $8.3 \times 10^{13}$& &or - $8 \times 10^{20}$ & \\ 
$136.875^{\circ}$ &$1.500^{\circ}$ &$0.102 \pm 0.003$ &$0.121 \pm 0.003$ &$4.12 \pm 0.1$ &$1.42 \times 10^{14} + 1.8 \times 10^{13}$ &$5.5 \pm 0.2$ &$1.35 \times 10^{21} + 2.5 \times 10^{20}$ &$1.1 \times 10^{21} \pm 3 \times 10^{20}$ \\ 
& & & & &or - $1.1 \times 10^{13}$& &or - $2.1 \times 10^{20}$ & \\ 
$137.000^{\circ}$ &$0.750^{\circ}$ & ... & ... &$8.1$ + $1.0$&$(< 4 \times 10^{12})$ & ... &$(7 \times 10^{19})$ &$< 2 \times 10^{20}$ \\
 & & & &or - $1.1$ & & & & \\
$137.000^{\circ}$ &$0.875^{\circ}$ & ... & ... &$8.9$ + $1.0$&$(< 4 \times 10^{12})$ & ... &$(< 7 \times 10^{19}$) &$< 2 \times 10^{20}$ \\
 & & & &or - $0.5$ & & & & \\
$137.000^{\circ}$ &$1.000^{\circ}$ &$\tau = 0.0231 \pm 0.002$ &$\tau = 0.0364 \pm 0.002$&$8.8$ + $0.8$ &$(4.9 \times 10^{13})$ &$3.8 \pm 0.2$ &$(4.7 \times 10^{20})$ &$7.6 \times 10^{20} \pm 2.3 \times 10^{20}$ \\ 
& & & &or - $0.9$ & & & & \\ 
$137.000^{\circ}$ &$1.125^{\circ}$ &$\tau = 0.0464 \pm 0.003$ &$\tau = 0.0113 \pm 0.002$ &$7.4$ + $1.6$ &$(6.2 \times 10^{13})$ &$26.6 \pm 0.2$ &$(5.9 \times 10^{20})$ &$5.3 \times 10^{21} \pm 1.6 \times 10^{21}$ \\ 
& & & &or - $0.6$ & & & & \\ 
$137.000^{\circ}$ &$1.250^{\circ}$ &$0.079 \pm 0.003$ &$\tau = 0.0108 \pm 0.002$ &$5.7 + 0.3$&$1.13 \times 10^{15}$ + $1.9 \times 10^{15}$ &$28.7 \pm 0.2$ &$1.08 \times 10^{22}$ + $1.8 \times 10^{22}$ &$5.7 \times 10^{21} \pm 1.7 \times 10^{21}$ \\ 
& & & &or - $0.2$ &or - $7.8 \times 10^{14}$ & &or - $7.6 \times 10^{21}$ & \\ 
$137.000^{\circ}$ &$1.375^{\circ}$ &$0.27 \pm 0.003$ &$0.0309 \pm 0.002$ &$5.2$ + $0.2$ &$9.79 \times 10^{14}$ + $6.4 \times 10^{14}$ &$17.8 \pm 0.2$ &$9.32 \times 10^{21}$ + $6.2 \times 10^{21}$ &$3.6 \times 10^{21} \pm 1.1 \times 10^{21}$ \\ 
& & &and $\tau = 0.01 \pm 0.002$ &or - $0.08$ &or - $2.3 \times 10^{14}$ & &or - $2.5 \times 10^{21}$ & \\ 
$137.000^{\circ}$ &$1.500^{\circ}$ &$0.15 \pm 0.003$ & ... &$5.0$ + $0.2$ &$4.19 \times 10^{14}$ + $1.6 \times 10^{14}$ &$11.6 \pm 0.1$ &$3.99 \times 10^{21}$ + $1.6 \times 10^{21}$ &$2.3 \times 10^{21} \pm 7 \times 10^{20}$ \\ 
& & & &or - $0.1$ &or - $8.2 \times 10^{13}$ & &or - $9.5 \times 10^{20}$ & \\ 
$137.125^{\circ}$ &$1.000^{\circ}$ & ... & ... &$7.1$ + $0.9$&$(< 4 \times 10^{12})$ & ... &$(<7 \times 10^{19})$ &$< 2 \times 10^{20}$ \\
 & & & &or - $0.4$ & & & & \\
$137.125^{\circ}$ &$1.125^{\circ}$ & ... & ... &$6.5$ + $0.5$&$(< 4 \times 10^{12})$ & ... &$(< 4 \times 10^{19})$ &$< 2 \times 10^{20}$ \\
 & & & &or - $0.2$ & & & & \\
$137.125^{\circ}$ &$1.250^{\circ}$ & $0.039 \pm 0.003$&$\tau = 0.005 \pm 0.002$ &$6.0$ + $0.2$& not foreground &$16.7 \pm 0.1$ & not foreground &$ 3.3 \times 10^{21} \pm 1 \times 10^{21}$ \\ 
& & & &or - $0.3$ & & & & \\ 
$137.125^{\circ}$ &$1.375^{\circ}$ &$0.29 \pm 0.003$ & $0.026 \pm 0.002$ &$5.3$ + $0.2$ &$1.24 \times 10^{15}$ + $1.2 \times 10^{15}$ &$12.5 \pm 0.2$ &$1.18 \times 10^{22}$ + $1.1 \times 10^{22}$ &$2.5 \times 10^{21} \pm 8 \times 10^{20}$ \\ 
& & &and $\tau = 0.001 \pm 0.001$ &or - $0.1$ &or - $3.5 \times 10^{14}$ & &or - $3.7 \times 10^{21}$ & \\ 
$137.125^{\circ}$ &$1.500^{\circ}$ &$0.067 \pm 0.003$ & ... &$5.2$ + $0.2$&$2.43 \times 10^{14}$ + $1.5 \times 10^{14}$ &$6.2 \pm 0.1$ &$2.31 \times 10^{21}$ + $1.5 \times 10^{21}$ &$1.2 \times 10^{21} \pm 4 \times 10^{20}$ \\ 
& & & &or - $0.2$ &or - $6.3 \times 10^{13}$ & &or - $6.8 \times 10^{20}$ & \\ 
$137.250^{\circ}$ &$1.000^{\circ}$ & ... & ... &$5.9$ + $8.6$&$(< 4 \times 10^{12})$ & ... &$(4 \times 10^{19})$ &$< 2 \times 10^{20}$ \\
 & & & &or - $0.3$ & & & & \\
$137.250^{\circ}$ &$1.125^{\circ}$ & ... & ... &$6.7$ + $1.6$&$(< 4 \times 10^{12})$ & ... & $(< 4 \times 10^{19})$ &$< 2 \times 10^{20}$ \\
 & & &and $0.003 \pm 0.002$ &or - $0.6$ & & & & \\
$137.250^{\circ}$ &$1.250^{\circ}$ & ... &$\tau = 0.009 \pm 0.002$&$7.7$ + $1.2$& not foreground &$2.5 \pm 0.1$ & not foreground &$5.0 \times 10^{20} \pm 2 \times 10^{20}$ \\ 
 & & & &or 0 $0.9$ & & & & \\ 
$137.250^{\circ}$ &$1.375^{\circ}$ & ... &$\tau = 0.029 \pm 0.002$ & $6.9$ + $0.7$& not foreground &$6.0 \pm 0.2$ & not foreground &$1.2 \times 10^{21} \pm 3 \times 10^{20}$ \\ 
 & & & &or - $0.7$ & & & & \\ 
$137.250^{\circ}$ &$1.500^{\circ}$ &$0.041 \pm 0.003$ & ... &$5.8 \pm 0.4$&$1.4 \times 10^{15} \pm 2.6 \times 10^{15}$ &$5.3 \pm 0.2$ &$1.3 \times 10^{22} \pm 2.5 \times 10^{22}$ &$1.1 \times 10^{21} \pm 3 \times 10^{20}$ \\
 & & & & & & & & \\
$137.375^{\circ}$ &$1.000^{\circ}$ & ... & ... &$5.6$ + $0.1$ &$(< 4 \times 10^{12})$ & ... &$(< 4 \times 10^{19})$ &$< 2 \times 10^{20}$ \\
 & & & &or - $0.2$ & & & & \\
$137.375^{\circ}$ &$1.125^{\circ}$ & ... & ... &$6.2$ + $1.9$&$(< 4 \times 10^{12})$ & ... &$(< 4 \times 10^{19})$ &$< 2 \times 10^{20}$ \\
 & & & &or - $0.4$ & & & & \\
$137.375^{\circ}$ &$1.250^{\circ}$ & ... & ... &$7.0$ + $0.9$&$(< 4 \times 10^{12})$ & ... &$(<7 \times 10^{19})$ &$< 2 \times 10^{20}$ \\
 & & & &or - $0.4$ & & & & \\
$137.375^{\circ}$ &$1.375^{\circ}$ & ... & ... &$7.1$ + $2.7$&$(< 4 \times 10^{12})$ & ... &$(< 4 \times 10^{19})$ &$< 2 \times 10^{20}$ \\
 & & & &or - $0.4$ & & & & \\
$137.375^{\circ}$ &$1.500^{\circ}$ & ... & ... &$6.1$ + $0.3$&$(< 4 \times 10^{12})$ & ... &$(< 4 \times 10^{19})$ &$< 2 \times 10^{20}$ \\
 & & & &or - $0.2$ & & & & \\
$137.500^{\circ}$ &$0.625^{\circ}$ &$\tau = 0.0034 \pm 0.002$ &$\tau = 0.0031 \pm 0.002$ &$6.0$ + $1.2$ &$(5.4 \times 10^{12})$ &$3.7 \pm 0.2$ &$(5.2 \times 10^{19})$ &$7.4 \times 10^{20} \pm 2.3 \times 10^{20}$  \\ 
 & & & &or - $0.4$ & & & & \\ 
$137.500^{\circ}$ &$0.750^{\circ}$ & ... & ... &$5.0$ + $0.04$& $< 1 \times 10^{13}$ & ... &$< 1 \times 10^{20}$ &$< 2 \times 10^{20}$ \\
 & & & &or - $0.2$ & & & & \\
$137.500^{\circ}$ &$0.875^{\circ}$ & ... & ... &$5.4$ + $0.3$& no upper limit & ... & no upper limit &$< 2 \times 10^{20}$ \\
 & & & &or - $0.2$ & & & & \\
$137.500^{\circ}$ &$1.000^{\circ}$ & ... & ... &$5.4 \pm 0.2$& $< 6 \times 10^{13}$ & ... &$< 5 \times 10^{20}$ &$< 2 \times 10^{20}$ \\
 & & & & & & & & \\
$137.500^{\circ}$ &$1.125^{\circ}$ & ... & ... &$5.7 \pm 0.2$&$(< 4 \times 10^{12})$ & ... &$(< 4 \times 10^{19})$ &$< 2 \times 10^{20}$ \\
 & & & & & & & & \\
$137.500^{\circ}$ &$1.250^{\circ}$ & ... & ... &$6.7$ + $1.7$&$(< 4 \times 10^{12})$ & ... &$(< 4 \times 10^{19})$ &$< 2 \times 10^{20}$ \\
 & & & &or - $0.4$ & & & & \\
$137.500^{\circ}$ &$1.375^{\circ}$ &$\tau = 0.0061 \pm 0.002$ &$\tau = 0.0131 \pm 0.002$ &$7.8$ + $1.3$&$(1.5 \times 10^{13})$ &$2.6 \pm 0.1$ &$(1.4 \times 10^{20})$ &$5.2 \times 10^{20} \pm 1.6 \times 10^{20}$ \\ 
 & & & &or - $0.7$ & & & & \\ 
$137.500^{\circ}$ &$1.500^{\circ}$ & ... & ... &$6.5$ + $0.6$&$(< 4 \times 10^{12}$) & ... &$(< 4 \times 10^{19})$ &$< 2 \times 10^{20}$ \\
 & & & &or - $0.4$ & & & & \\
$137.625^{\circ}$ &$1.000^{\circ}$ & ... & ... &$5.6$ + $1.3$&$(< 4 \times 10^{12})$ & ... &$(< 4 \times 10^{19})$ &$< 2 \times 10^{20}$ \\
 & & & &or - $0.3$ & & & & \\
$137.625^{\circ}$ &$1.125^{\circ}$ & ... & ... &$5.9$ + $0.1$&$(< 4 \times 10^{12})$  & ... &$(< 4 \times 10^{19})$ &$< 2 \times 10^{20}$ \\
 & & & &or - $0.2$ & & & & \\
$137.625^{\circ}$ &$1.250^{\circ}$ & ... & $\tau = 0.003 \pm 0.002$ &$6.9$ + $0.4$&$(3.5 \times 10^{12})$&$4.5 \pm 0.2$ &$(3.4 \times 10^{19})$ &$9.0 \times 10^{20} \pm 3 \times 10^{20}$ \\
 & & & &or - $0.5$ & & & & \\
$137.625^{\circ}$ &$1.375^{\circ}$ & ... & $\tau = 0.003 \pm 0.002$ &$7.2$ + $0.9$& $(3.5 \times 10^{12})$ &$26.9 \pm 0.2$ & $(3.4 \times 10^{19})$ &$5.4 \times 10^{21} \pm 2 \times 10^{21}$ \\
 & & & &or - $0.7$ & & & & \\
$137.625^{\circ}$ &$1.500^{\circ}$ &$0.026 \pm 0.003$ &$0.5 \pm 0.003$ &$6.0$ + $0.5$&not foreground &$17.0 \pm 0.2$ &not foreground &$3.4 \times 10^{21} \pm 1 \times 10^{21}$ \\
 & & & &or - $0.6$ & & & & \\
$137.750^{\circ}$ &$1.125^{\circ}$ & ... & ... &$6.1$ + $0.4$&$(< 4 \times 10^{12})$ & ... &$(< 4 \times 10^{19})$ &$< 2 \times 10^{20}$ \\
 & & & &or - $0.3$ & & & & \\
$137.750^{\circ}$ &$1.250^{\circ}$ & ... & ... &$7.0$ + $0.4$&$(< 4 \times 10^{12})$ & ... &$(< 4 \times 10^{19})$ &$< 2 \times 10^{20}$ \\
 & & & &or - $0.3$ & & & & \\
$137.750^{\circ}$ &$1.375^{\circ}$ &$0.027 \pm 0.003$ &$0.063 \pm 0.003$ &$7.0$ + $0.6$&not foreground &$23.5 \pm 0.2$ &not foreground &$4.7 \times 10^{21} \pm 1.4 \times 10^{21}$ \\
 & & & &or - $0.5$ & & & & \\
$137.750^{\circ}$ &$1.500^{\circ}$ &$\tau = 0.003 \pm 0.002$ &$0.007 \pm 0.002$ &$7.8$ + $1.2$&$(7.3 \times 10^{12})$ &$28.4 \pm 0.2$ &$(6.9 \times 10^{19})$ &$5.7 \times 10^{21} \pm 1.7 \times 10^{21}$ \\ 
& & &and $\tau = 0.002 \pm 0.001$ &or - $0.8$ & & & & \\ 
$137.875^{\circ}$ &$1.250^{\circ}$ & ... & ... &$6.8$ + $1.8$&$(< 4 \times 10^{12})$ & ... &$(< 4 \times 10^{19})$ &$< 2 \times 10^{20}$ \\
 & & & &or - $0.1$ & & & & \\
$137.875^{\circ}$ &$1.375^{\circ}$ & ... & ... &$7.3$ + $0.2$&$(< 4 \times 10^{12}$) & ... &$(< 4 \times 10^{19})$ &$< 2 \times 10^{20}$ \\
 & & & &or - $0.3$ & & & & \\
$137.875^{\circ}$ &$1.500^{\circ}$ & ... & ... &$7.3$ + $0.7$&$(< 4 \times 10^{12})$ & ... &$(< 4 \times 10^{19})$ &$< 2 \times 10^{20}$ \\
 & & & &or - $0.4$ & & & & \\
$137.875^{\circ}$ &$1.625^{\circ}$ &$\tau = 0.0063 \pm 0.002$ &$\tau = 0.0124 \pm 0.002$ &$7.1$ + $0.4$&$(1.4 \times 10^{13})$ &$9.3 \pm 0.1$ &$(1.4 \times 10^{20})$ &$1.9 \times 10^{20} \pm 6 \times 10^{20}$ \\ 
 & & & &or - $0.2$ & & & & \\ 
$137.875^{\circ}$ &$1.750^{\circ}$ & ... & ... &$5.6$ + $0.3$& no upper limit & ... & no upper limit &$< 2 \times 10^{20}$ \\
 & & & &or - $0.5$ & & & & \\
$138.000^{\circ}$ &$1.250^{\circ}$ & ... & ... &$5.9 \pm 0.3$&$(< 4 \times 10^{12})$ & ... &$(< 4 \times 10^{19})$ &$< 2 \times 10^{20}$ \\
 & & & & & & & & \\
$138.000^{\circ}$ &$1.375^{\circ}$ & ... & ... &$6.5$ + $0.6$&$(< 4 \times 10^{12})$ & ... &$(< 4 \times 10^{19})$ &$< 2 \times 10^{20}$ \\
 & & & &or - $0.2$ & & & & \\
$138.000^{\circ}$ &$1.500^{\circ}$ & ... & ... &$7.2$ + $0.8$&$(< 4 \times 10^{12})$ & ... &$(< 4 \times 10^{19})$ &$< 2 \times 10^{20}$ \\
 & & & &or - $0.5$ & & & & \\
$138.000^{\circ}$ &$1.625^{\circ}$ & ... &$\tau = 0.0024 \pm 0.002$ &$7.7$ + $4.8$&$(3.6 \times 10^{12})$ &$1.4 \pm 0.1$ &$(3.4 \times 10^{19})$ &$2.8 \times 10^{20} \pm 9 \times 10^{19}$ \\ 
 & & & &or - $0.5$ & & & & \\ 
$138.000^{\circ}$ &$1.750^{\circ}$ & ... & $\tau = 0.006 \pm 0.002$ &$6.0$ + $0.3$& $(7.1 \times 10^{12})$ &$6.7 \pm 0.1$ & $(6.8 \times 10^{19})$ &$1.3 \times 10^{21} \pm 4.0 \times 10^{20}$ \\
 & & & &or - $0.4$ & & & & \\
$138.125^{\circ}$ &$1.250^{\circ}$ & ... & ... &$5.5$ + $0.7$& no upper limit & ... & no upper limit &$< 2 \times 10^{20}$ \\
 & & & &or - $0.2$ & & & & \\
$138.125^{\circ}$ &$1.375^{\circ}$ & ... & ... &$5.9$ + $0.5$&$(< 4 \times 10^{12})$ & ... &$(< 4 \times 10^{19})$ &$< 2 \times 10^{20}$ \\
 & & & &or - $0.3$ & & & & \\
$138.125^{\circ}$ &$1.500^{\circ}$ & ... & ... &$6.9$ + $0.5$&$(< 4 \times 10^{12})$ & ... &$(< 4 \times 10^{19})$ &$< 2 \times 10^{20}$ \\
 & & & &or - $0.2$ & & & & \\
$138.125^{\circ}$ &$1.625^{\circ}$ & ... & ... &$6.7 \pm 0.4$&$(< 4 \times 10^{12})$ & ... &$(< 4 \times 10^{19})$ &$< 2 \times 10^{20}$ \\
 & & & & & & & & \\
$138.125^{\circ}$ &$1.750^{\circ}$ & ... & ... &$5.6$ + $0.8$& no upper limit &$6.8 \pm 0.1$ & no upper limit &$1.4 \times 10^{21} \pm 4 \times 10^{20}$ \\
 & & & &or - $0.3$ & & & & \\
$138.250^{\circ}$ &$1.250^{\circ}$ & ... &$0.011 \pm 0.003$ &$5.3$ + $0.3$&not foreground &$0.5 \pm 0.1$ &not foreground &$1.0 \times 10^{20} \pm 3.6 \times 10^{19}$ \\
 & & & &or - $0.2$ & & & & \\
$138.250^{\circ}$ &$1.375^{\circ}$ & ... & ... &$5.8 \pm 0.2$&$(< 4 \times 10^{12}$) & ... &$(< 4 \times 10^{19}$) &$< 2 \times 10^{20}$ \\
 & & & & & & & & \\
$138.250^{\circ}$ &$1.500^{\circ}$ & & &$6.3$ + $0.8$& &$7.5 \pm 0.2$ & &$1.5 \times 10^{21} \pm 5 \times 10^{20}$ \\ 
 & & & &or - $0.2$ & & & & \\ 
$138.250^{\circ}$ &$1.625^{\circ}$ & ... & ... &$6.3$ + $1.5$&$(< 4 \times 10^{12})$ &$6.9 \pm 0.1$ &$(< 4 \times 10^{19}$) &$1.4 \times 10^{21} \pm 4 \times 10^{20}$ \\
 & & & &or - $0.3$ & & & & \\
$138.250^{\circ}$ &$1.750^{\circ}$ &$0.01 \pm 0.003$ &$0.014 \pm 0.003$ &$5.2$ + $1.0$&not foreground &$10.7 \pm 0.1$ &not foreground &$2.1 \times 10^{21} \pm 6 \times 10^{20}$ \\
 & & & &or - $0.3$ & & & & \\
\enddata
\end{deluxetable*}
\end{longrotatetable}
\end{center}




\clearpage

\end{document}